# Suppression of Droplet Breakage by Early Onset of Interfacial Instability

Rutvik Lathia, Chandantaru Dey Modak, and Prosenjit Sen*

Centre for Nano Science and Engineering, Indian Institute of Science, Bangalore, India, 560012

***Corresponding Author's Email:** prosenjits@iisc.ac.in

## Abstract:

### Hypothesis:

Interfacial instabilities cause undesirable droplet breakage during impact. Such breakage affects many applications, such as printing, spraying, etc. Particle coating over a droplet can significantly change the impact process and stabilize it against breakage. This work investigates the impact dynamics of particle-coated droplets, which mostly remains unexplored.

### Experiments:

Particle-coated droplets of different mass loading were formed using a volume addition. Then the prepared droplets were impacted on superhydrophobic surfaces, and their dynamics were recorded using a high-speed camera.

### Findings:

We report an intriguing phenomenon where interfacial fingering instability helps suppress breakage in particle-coated droplets. This island of breakage suppression, where the droplet maintains its intactness upon impact, appears within a regime of Weber numbers where droplet breakage is inevitable. The onset of fingering instability in particle-coated droplets is observed at much lower impact energy, around two times less than the bare droplet. The instability is characterized using the rim Bond number. The instability suppresses breakage because of the higher losses associated with the formation of stable fingers. Such instability can also be seen in Leidenfrost surfaces and dust/pollen-covered surfaces, making it useful in many applications related to self-cleaning.

# 1 Introduction

Droplet breakage on impact is undesirable. It affects essential applications such as printing, spraying (e.g., pesticides), coating, bioreactors, cooling, and directional transport [1–3]. In printing, droplet breakage leads to the generation of undesired spots and compromises printing resolution. In the case of spraying on plants, ejected droplets are lost to the ground and cause environmental pollution [4]. Similar droplet breakage also causes pollution during fertilizer production [3]. Smaller droplets are more efficient in spreading viruses and diseases than larger ones. Thus, smaller droplet generation due to impact breakage enhances the spreading of pathogens and diseases in plants [5,6] and humans [7] onto a larger area.

Droplet impact on surfaces is studied for their implications in various applications and natural phenomena. On impact, the droplet spreads, and its kinetic energy is stored as surface energy. It converts back into kinetic energy during rebound. While the top part of the droplet rapidly moves away from the surface, the bottom part leaves the surface slowly as adhesion delays detachment [8]. This process results in the stretching of the droplet with its stretched length ($L_{max}$) larger than the maximum spread diameter ($D_{max}$). Stretching increases with impact velocity. Beyond a critical Weber number, the stretching is sufficient to enable the ejection of smaller droplets from the top. Weber number is defined as $We = \rho V^2 D_0/\gamma$ , where $\rho, V, D_0$ and $\gamma$ are density, impact velocity, diameter, and surface tension, respectively. This is driven by Rayleigh-Plateau (RP) instability. As the impact velocity increases, the stretched interface breaks down into several droplets during the lift-off phase [9]. This kind of dissociation is referred to as pinch-off. At higher impact velocities, droplet dissociates on the surface due to the onset of finger formation driven by Rayleigh-Taylor (RT) and Rayleigh-Plateau (RP) instability [10–12]. This type of dissociation is referred to as a receding breakup.

We report suppression of droplet breakage with the help of hydrophobic particle coating. Particle-coated droplets are commonly known as liquid marbles (LM). The impact of LM on the superhydrophobic surface reveals a higher Weber number stability regime which is non-existent in the bare droplet impact. Interestingly for LM, as the impact velocity increases, droplet breakage does not progress from pinch-off to receding breakup regime. The pinch-off regime is shortened, and we observe complete suppression of droplet breakage in LM for a range of higher impact velocities.

This study presents a technique for breakage suppression, which will be helpful in many applications, such as in the continuous production of mechanically stable bioreactors [13,14]. LM has also been used as a biological model [15]. Studying its large-deformation dynamics helps us understand the response of organs and cells under sudden impact conditions such as accidents [16]. Besides, the present study also helps us understand the dynamics of particle-coated curved interfaces at large deformations, which are largely unexplored. As the applications of LM in the digital microfluidics platform are increasing [17], the present paper also helps design such applications better. A similar phenomenon is observed for bare droplet impact on particle-covered surfaces and LM impact on Leidenfrost surfaces. Such scenarios help in understanding the interactions of the droplet with particles for the applications like self-cleaning and pollen dispersion [6,18–20].

# 2 Experimental methods

## 2.1 Preparation of liquid marble

A polytetrafluoroethylene (PTFE) powder with an average particle diameter of 35 μm was used to prepare liquid marble (with a core liquid as DI water of 8.2 μL) by the rolling method [21].

Control of the mass loading was obtained by creating smaller liquid marble (LM) of different volumes with full surface coverage. Then the LM is placed on a superhydrophobic surface with a bare droplet. Collision of the LM with a bare droplet results in merging and forming a larger LM, as shown in Fig. 1(A). The final LM volume was 8.2 μL. Superhydrophobic surface is required to avoid rupturing of LM with lower mass loading. The geometric relation $S^3 \sim V^2$ determines the initial LM volume, where $S$ and $V$ are the surface area and the volume of the liquid drop, respectively [24]. The different volumes of LMs and water droplets used to prepare various mass loading are described in Table 1. Where $V_{LM}$ and $V_W$ represent the volume of liquid marble and the volume of water drops. The mass loading was determined by averaging the mass of ten liquid marbles after completely drying the liquid.

## 2.2 Preparation of superhydrophobic surface

Superhydrophobic copper surface was prepared by the method reported previously [25,26]. A copper substrate (3 cm × 2 cm) was cleaned with acetone, isopropyl alcohol (IPA), and deionized (DI) water. This was followed by a 30 s cleaning with sulfuric acid (33% in DI water). The cleaned copper surface was then immersed in an aqueous solution of sodium hydroxide (2.5 mol/L) and ammonium persulfate (0.1 mol/L) for 20 min at room temperature. This solution etches the copper surface and produces copper hydroxide nanowires on the surface (Supplementary Fig. S1). The substrate is cleaned multiple times with DI water and dried with nitrogen. The substrate is dipped in Teflon solution for 10 min to get a superhydrophobic surface. Subsequently, it is dried by heating at 110 °C for 10 min. The prepared superhydrophobic surfaces show excellent repellency with a water contact angle of 171° (Supplementary Fig. S1).

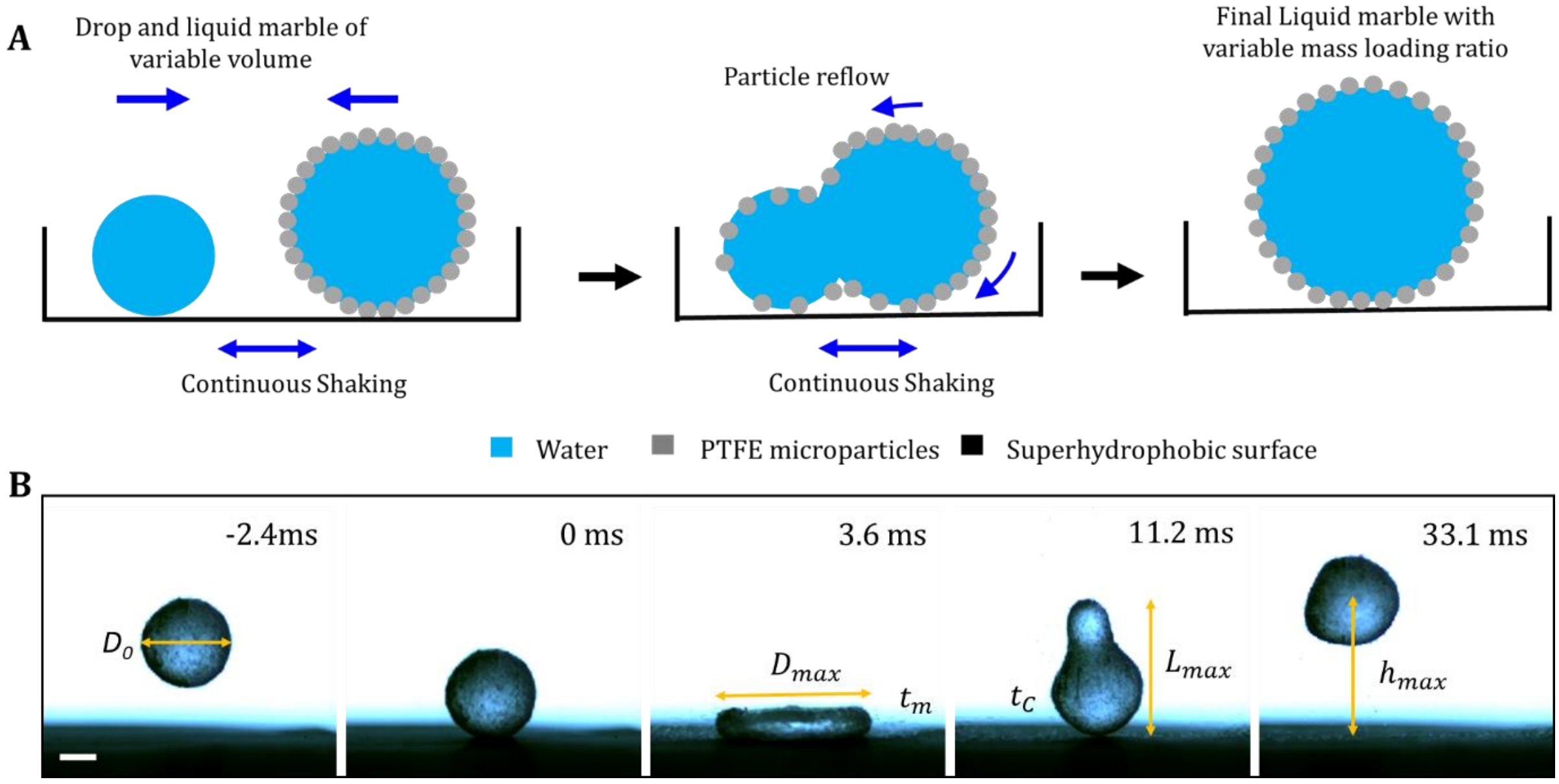

**Fig. 1.** (A) Methodology for the preparation of LM with variable mass loading. First, a smaller volume LM with maximum mass loading is prepared. The LM volume is increased by merging with a bare droplet. The droplet volume is taken such that the final volume of the LM will be 8.2 μL. Vibration is necessary to ensure uniform particle distribution across the surface. (B) Parameters measured during LM impact. Where $D_0$, $D_{max}$, $t_m$, $t_c$, $L_{max}$, and $h_{max}$, are the initial diameter of the LM, maximum spread diameter, maximum spreading time, contact time, maximum length during rebound, and the maximum height of rebound, respectively. (Scale bar – 1 mm)

| $V_{LM}$ (μL) | $V_w$ (μL) | ML (μg/mm²) |
|---|---|---|
| 0 | 8.2 | 0 |
| 1.58 | 6.42 | $5.87 \pm 0.49$ |
| 4.46 | 3.74 | $10.77 \pm 0.47$ |
| 8.2 | 0 | $16.63 \pm 0.5$ |

**Table 1**: The values of volume to be taken for particular mass loading. As described in the main text, various sizes of LM and droplets collision results in different mass loading. $V_{LM}$ and $V_W$ represent the volume of liquid marble and water droplet, respectively. The final volume of LM used in our experiments is fixed at 8.2 μL.

### 2.3 Droplet impact experiments

Prepared LM with different mass loading is impacted over a superhydrophobic surface. Impact velocity was controlled by impacting the LM from different heights, and the dynamics were recorded with a high-speed camera. The recorded images were analyzed for measuring different parameters, as shown in Fig. 1(B).

## 3 Results and discussion

### 3.1 The dynamic surface tension of LM during impact

For static scenarios, LM has reduced effective surface tension. As shown in Supplementary Fig. S1(A), the effective surface tension ($\gamma_{eff}$) is determined by the maximum puddle height method. The effective surface tension decreases with mass loading increase [22]. For lower mass loaded LM, $\gamma_{eff}$ stays near the bare droplet. Whereas, for higher mass loading $\gamma_{eff}$ decreases significantly to ~ 50 mN. However, in the case of droplet impact, the dynamic surface tension of the LM determines the interface dynamics. The value of dynamic surface tension lies between the liquid's surface tension and the effective surface tension ($\gamma_{eff}$) of LM. During spreading, the newly created surface is primarily a liquid surface with very few particles; thus, the change in net surface energy during the spreading process is determined by the liquid's surface tension. This scenario is analogous to the impact of droplets with slow diffusing surfactants [27,28]. Thus, as an approximation, water surface tension has been used as the dynamic surface tension in our case.

The spreading dynamics of an LM show approximately the same behavior and scaling as a bare droplet (Fig. 2). Maximum spreading diameter ($D_{max}$) on the superhydrophobic surface is known to follow $D_{max} \sim D_0\, We^{0.25}$ scaling [29]. LM follows approximately the same scaling $D_{max} \sim D_0\, We^{0.27}$ (Fig. 2(A)). As shown in Fig. 2(B), the maximum spread time ($t_m$) normalized with the drop capillary time ($\tau_d \sim \sqrt{\rho D_0^3/\gamma}$) is also similar. Both these observations justify the use of liquid surface tension as an approximation for the dynamic surface tension.

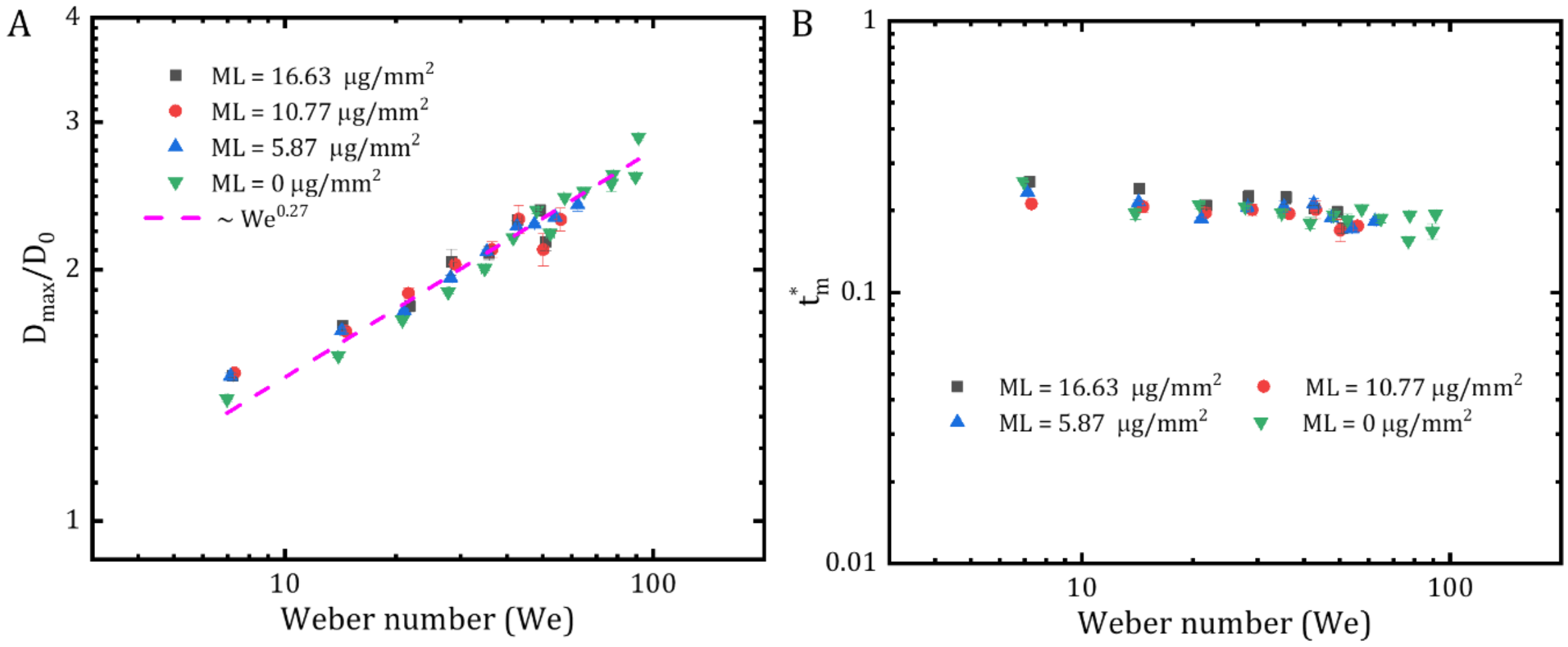

**Fig. 2.** (A) The normalized maximum diameter plotted against $We$. Power law fit leads to $We^{0.27}$. (B) The maximum spread time ($t_m$) normalized with drop capillary time ( $\tau = \sqrt{\rho D_0^3/\gamma}$ ).

## 3.2 Impact outcome

The rebound of the LM follows entirely different retraction dynamics, which leads to a different outcome for the impact process. With an increase in impact velocity for a bare droplet, the outcome progresses from a no-breakage to pinch-off to receding-breakup, as shown in Fig. 3. Interestingly, as the impact velocity increases, LM breakage does not progress from pinch-off to receding-breakup regime (Supplementary Video S1). The pinch-off regime is shortened, and we observe complete suppression of LM breakage for a range of higher impact velocities. In this zone, LM shows intactness and mechanical stability even at high $We$. Traditionally polymer additives are used to increase viscosity for breakage suppression [30]. However, it is impossible to explain our observations through arguments of increased viscosities as the island of "no-breakage" emerges between two "dissociation" regimes. Droplets with enhanced viscosity will only show the transition from no-breakage to breakage. Our study reveals that this suppression is induced by the early onset of fingering instability in LM.

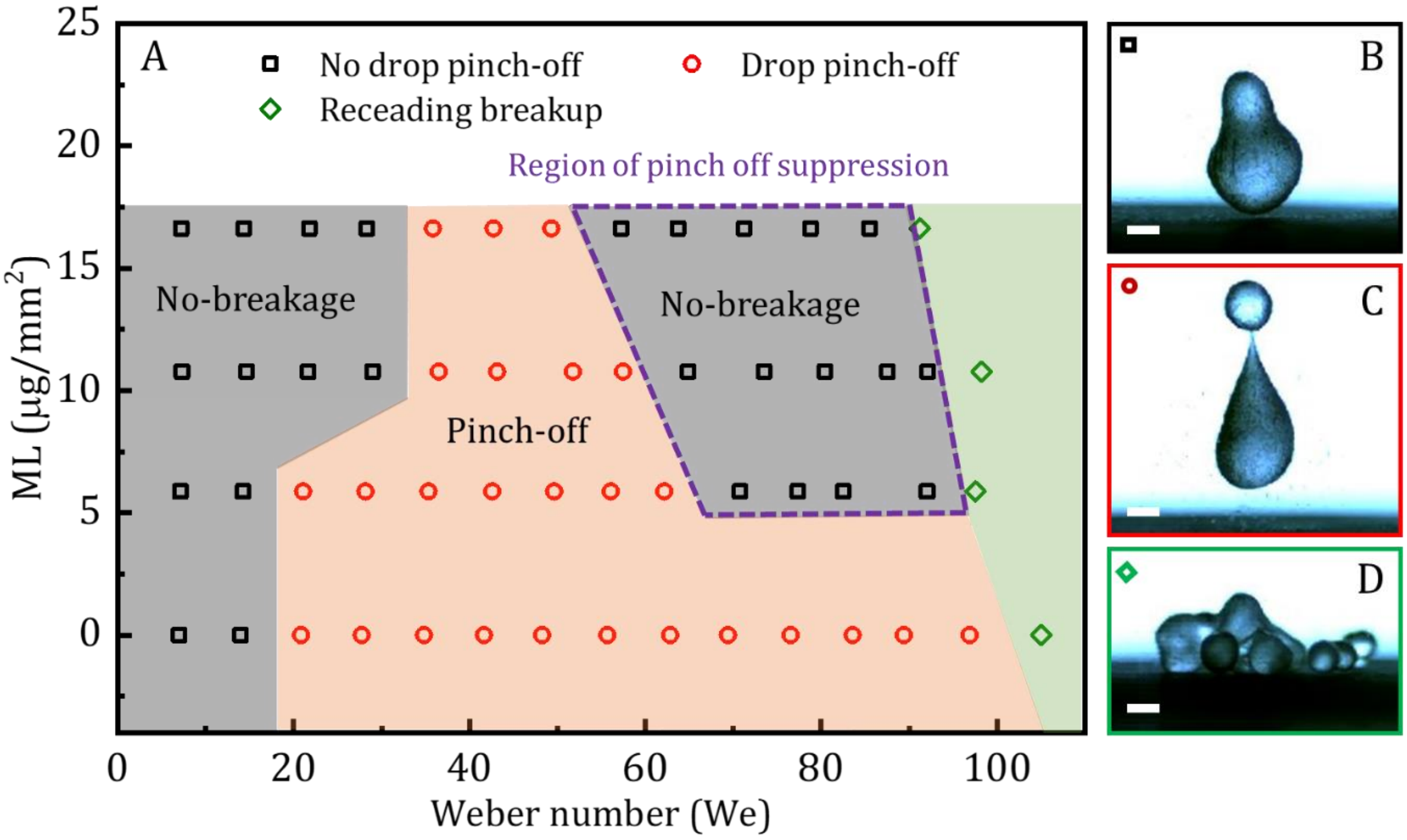

**Fig. 3.** (A) Droplet stability regime for different mass loading and impact *We*. Photographs of the different scenarios for mass loading = 5.87 μg/mm². (B) *No-breakage*: (black squares) represents the stable LM without any rupture or pinch-off. (C) *Pinch-off*: (red circles) droplet breaks from the top during rebound. (D) *Receding-breakup*: (green diamonds) multiple droplets eject on the substrate during the retraction phase. (Scale bar – 1 mm)

As the droplet rebounds and tries to leave the surface, it stretches. Stretching depends on the remaining kinetic energy in the rebounding droplet and the surface adhesion. Droplet pinch-off through Rayleigh-Plateau (RP) instability is only possible beyond a critical stretching ratio ($L_{max}/D_0$), where $L_{max}$ is the stretched length of the droplet during rebound (see, Inset Fig. 4(A)). As seen in Fig. 4(A), we identify the critical stretching ratio of $L_{max}/D_0 \sim 1.9$ in our experiments. As seen in Fig. 4(B), at the lowest $D_{max}$ ($We < 20$), the stretched length ($L_{max}$) is approximately equal to the maximum spread diameter ($D_{max}$) in all cases. This behavior indicates an inviscid impact with negligible surface adhesion. However, for large impact velocities ($We > 57$), we observe a continuous reduction in $L_{max}$ for high mass loading LM. $L_{max}/D_0$ values reduce below the critical stretching ratio, and droplet pinch-off is suppressed. In contrast, $L_{max}$ remains high for bare droplets. We attribute this reduction in $L_{max}$ and the associated recurrence of the no-breakage regime to an additional dissipation mechanism. This dissipation stems from the instability that leads to the formation of finger-like structures.

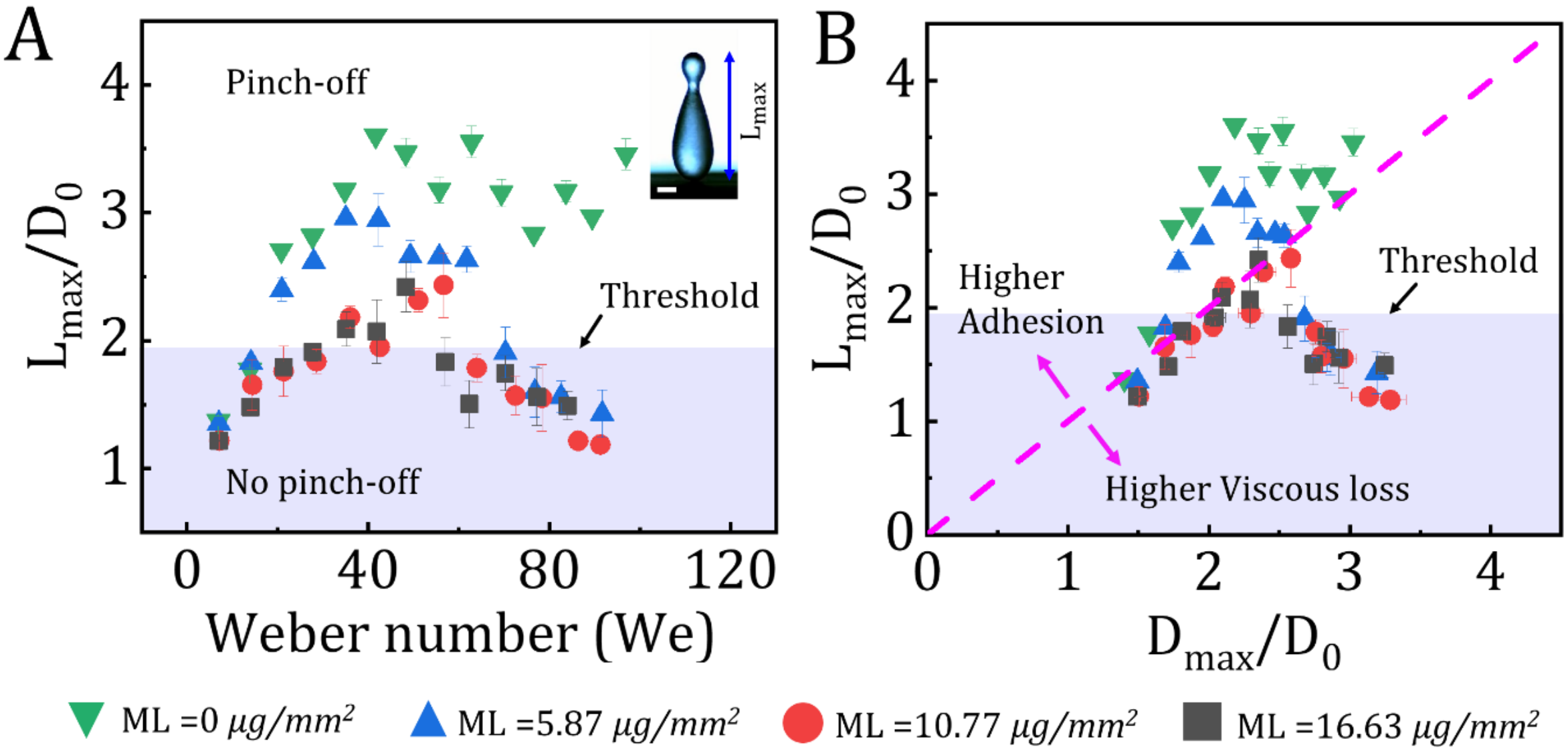

**Fig. 4.** (A) The normalized maximum extension is plotted against $We$. The colored region corresponds to $L_{max}/D_0 < 1.9$. Inset: representation of $L_{max}$ during rebound. (Scale bar - 1 mm). (B) The normalized maximum extension is plotted against the normalized maximum diameter. Here, the dashed line corresponds to $L_{max} \cong D_{max}$. The data above the line represents higher adhesion energy, and the data below represents a higher viscous loss.

### 3.3 Rim instability

As the droplet reaches its maximum spread diameter, a rim of liquid is formed at the edge of the flattened droplet. At $We > 57$, an interfacial instability sets in during the spreading phase. As shown in Fig. 5(A) & (B) and Supplementary Video S2, this rim destabilizes as it expands, and perturbations can be seen in both bare droplets and LM However, only for the LM case, these perturbations grow with time and form a finger-like structure during the retraction phase. Finger formation is not observed for the bare droplet case at the same $We$.

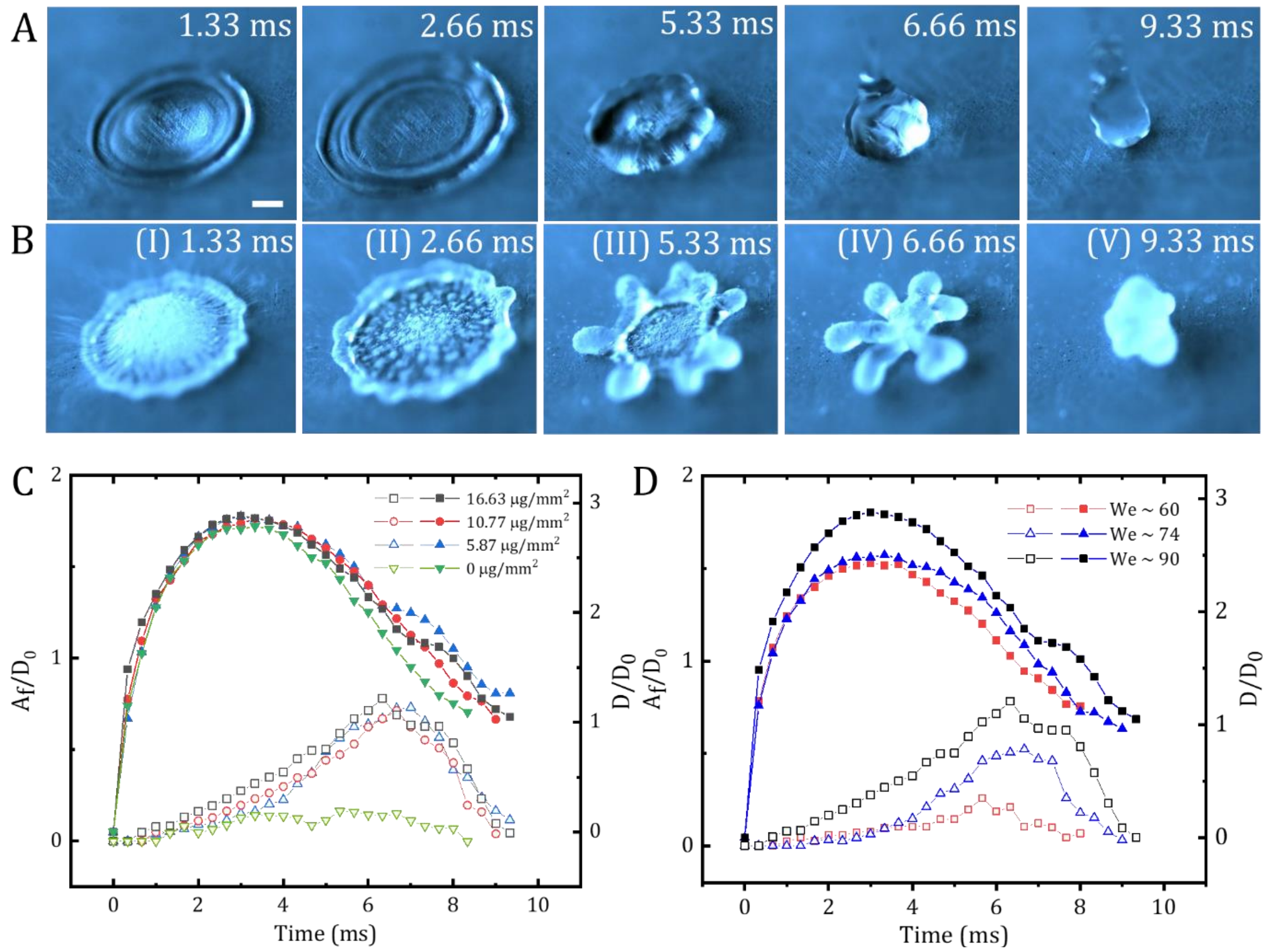

**Fig. 5.** (A) Sequential snapshots of a water droplet impacting the superhydrophobic surface for *We* ~ 77. (B) LM impact on the superhydrophobic surface at *We* ~ 77. Initial finger formation can be seen at the maximum spread. (C) Temporal evolution of normalized finger length ($A_f/D_0$) and diameter ($D/D_0$) for various mass loading at *We* ~ 90. The filled symbols represent the normalized diameter evolution ($D/D_0$), and the unfilled symbols represent the normalized finger length ($A_f/D_0$). Here, $A_f$ is the amplitude of the fingers. (D) Temporal evolution of normalized finger length and diameter for mass loading of 16.63 μg/mm$^2$ at different *We*.

Fig. 5(C) represents the temporal evolution of normalized finger length and spread diameter for $We$ ~ 90. The evolution of diameter is nearly the same for all LMs and the bare droplet. However, the finger length remains negligent in the bare droplet case. Additionally, the growth of the perturbation is not limited to the spreading phase only. The perturbations grow faster after the droplet reaches its maximum spread state (~ 2.66 ms). Compared to bare droplets, perturbation at maximum spread is higher in LM. The perturbation growth is further driven by the rapid retraction of the interface between the fingers while the tips remain nearly stationary. For a particular LM, an increase in $We$ results in increased spread diameter and finger length, as seen in Fig. 5(D).

The formation of fingers for a bare droplet is observed at a much higher Weber number (~110). However, the formation of fingers for bare droplets shows very distinct differences. In contrast to LM fingers, they dissociate on the surface due to receding breakup. Fig. 6(A) represents the variation of normalized finger amplitude ($A_f/D_0$) at the maximum spread. At $We$ below 110, the LM has much higher amplitude compared to the bare droplet. However, above $We \sim 110$, a similar trend is observed. Maximum finger length also follows the same trend, and LM shows large finger growth even at lower $We$ (Fig. 6(B)). In previous literature, the generation of fingers has been attributed to Rayleigh-plateau (RP), Rayleigh-Taylor (RT) and Kelvin-Helmholtz (KH) instabilities [11,31–33].

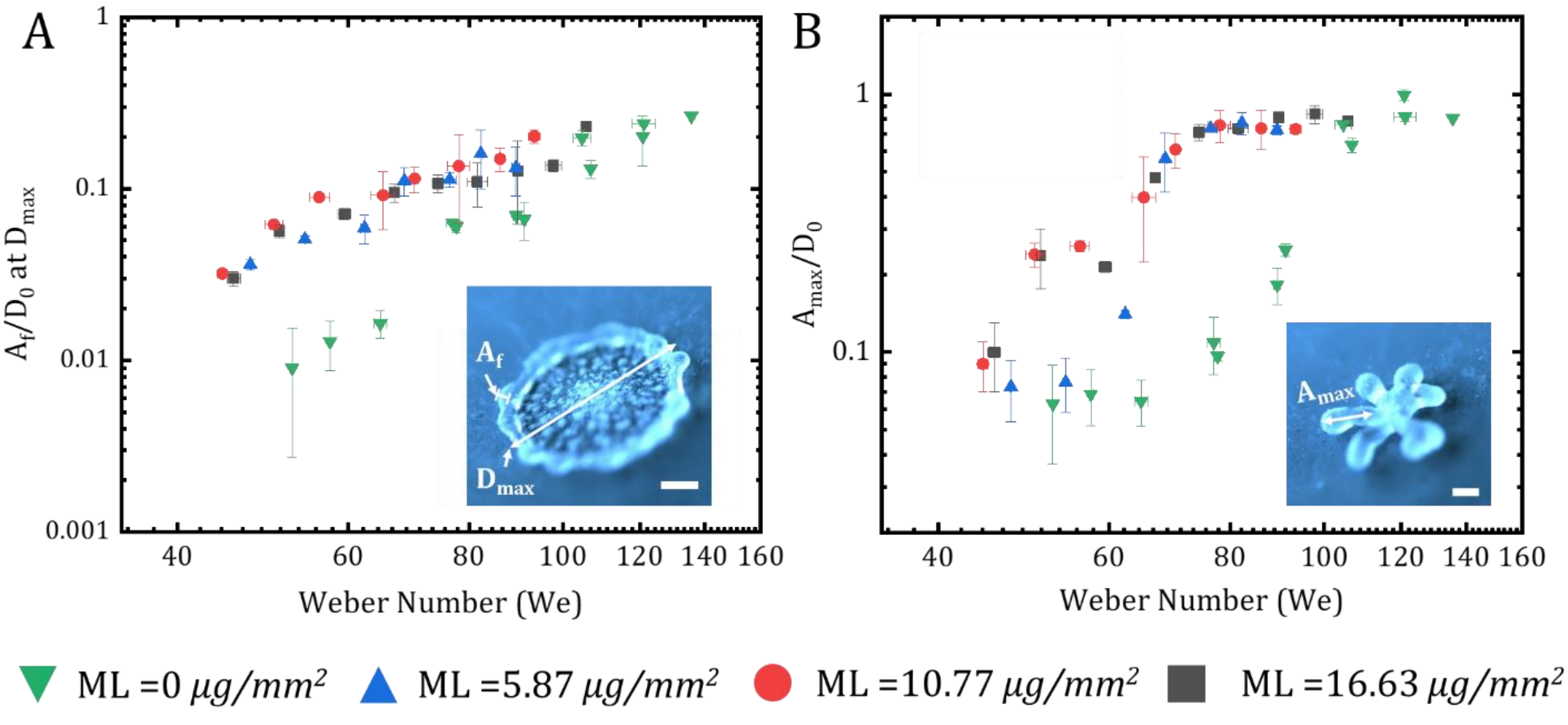

**Fig. 6.** (A) Variation of normalized finger length ($A_f/D_0$) at maximum spread diameter for different Weber numbers. Clear distinction can be seen between the bare droplet and LM Inset

shows the measurement of finger amplitude ($A_f$) and maximum diameter ($D_{max}$). (B) The normalized maximum amplitude of the finger ($A_{max}/D_0$) is plotted against the Weber number.

### 3.4 Rim Bond number

The coupled effect of RT and RP instability has been attributed to finger formation [12]. The work has analyzed the stability of an inviscid cylindrical rim subjected to acceleration. The dispersion relation for the coupled effect is given in terms of nondimensionalized wavenumber ($\kappa$) and rim Bond number ($Bo$) as [12]

$$\begin{aligned} \omega^2 &= 0.5\left\{-\chi(\kappa) + \sqrt{\chi(\kappa)^2 - 4\psi(\kappa)}\right\} \\ \chi(\kappa) &= \frac{\kappa I_1(\kappa)}{I_0(\kappa)}(\kappa^2 - 1) + \frac{\kappa I_2(\kappa)}{I_1(\kappa)}\kappa^2 \\ \psi(\kappa) &= \frac{\kappa^2 I_2(\kappa)}{2I_0(\kappa)}\left[2(\kappa^2 - 1)\kappa^2 - \left(\frac{Bo}{4}\right)^2\right] \end{aligned} \quad (1)$$

where, $\omega$ is the growth rate non-dimensionalized by rim capillary time ( $\tau_r = \sqrt{\rho b^3/8\gamma}$ ). $\kappa$ is perturbation wavenumber non-dimensionalized by rim radius ($b/2$). $I_n(\kappa)$ is the modified Bessel function of the first kind of order $n$. $Bo$ is the rim Bond number given by $\rho a b^2/\gamma$, where $a$ is the rim deceleration. Dispersion curves (Eq. (1)) for different rim Bond numbers are plotted in Supplementary Fig. S2. For the range of $Bo$ observed in our experiments, the curves remain close to that of the pure RP instability.

Finger length is determined by the growth rate of the fastest growing unstable mode. Supplementary Fig. S3 shows the theoretical growth rate for the fastest growing mode with $Bo$. Fig 7(A) plots the normalized finger length ($A_f$) at maximum spread with $Bo = \rho a b^2/\gamma$. Here, deceleration is approximated as $a \cong D_{max,i}/t_m^2$, where $D_{max,i}$ is the inner rim diameter at maximum spread as shown in Fig 7(A) inset. It is observed that the measured finger length is well explained with the rim Bond number. Fig. 7(B) plots the maximum finger length attained during the retraction phase as a function of the finger length at $D_{max}$. Supplementary Fig. S4 plots normalized $A_{max}$ with the rim Bond number. Below a critical $Bo$ (~ 4), where the initial finger length is high ($A_f/D_0$ > 0.07), the fingers grow during the retraction phase. At higher $Bo$ where $A_f/D_0$ < 0.07, the finger growth is restricted during the retraction phase.

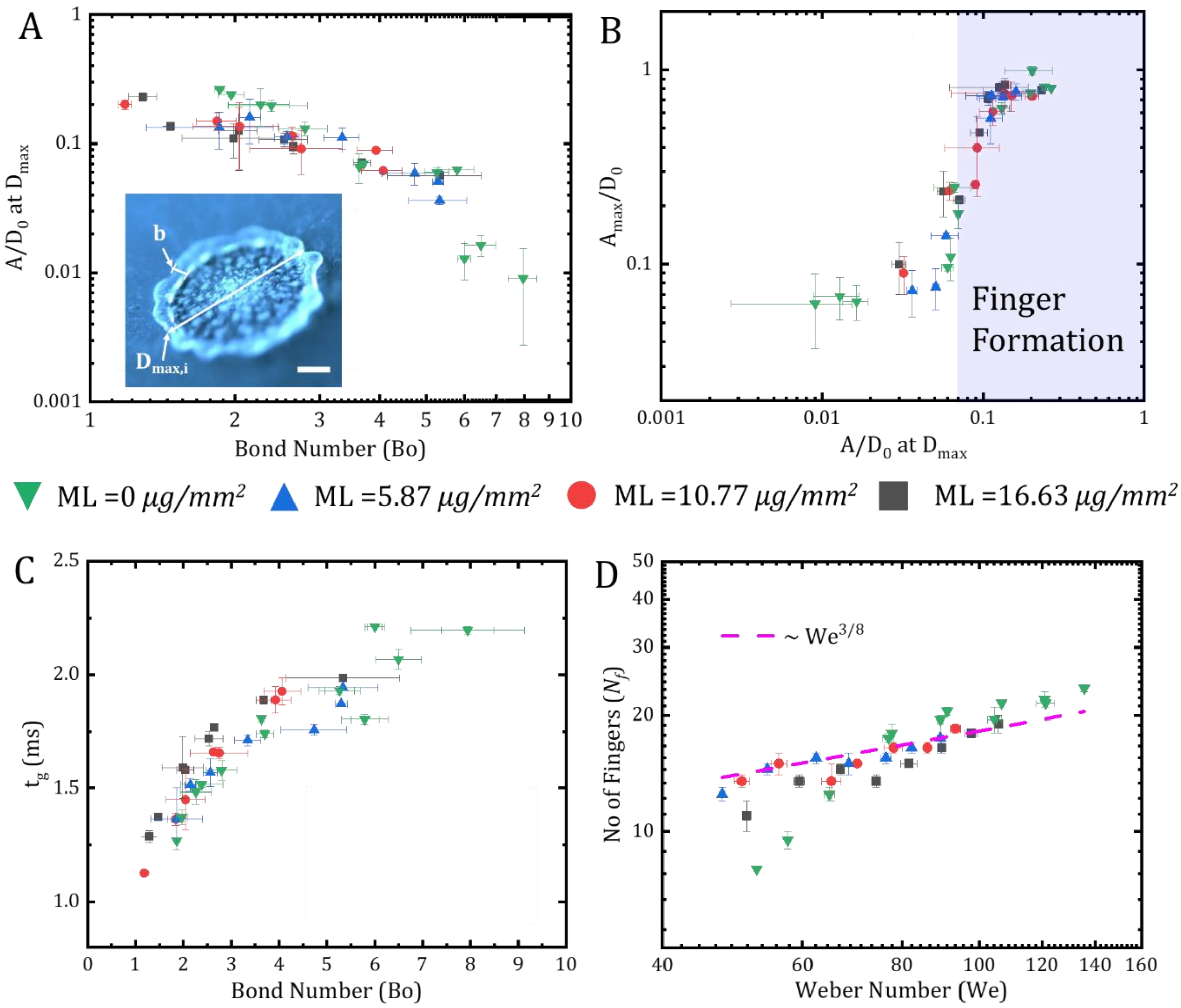

**Fig. 7.** (A) Variation of normalized finger length ($A_f/D_0$) at maximum spread diameter for different rim Bond numbers. Inset: the LM at maximum spread where, the rim width is given by $b$ and inner rim diameter is given by $D_{max,i}$.(B) The maximum amplitude of the fingers is plotted against the amplitude at the maximum spread. Above critical value of $A_f/D_0$ ~ 0.07, the finger formation is predominant. Additionally, all data collapse into one single curve, representing amplitude at maximum spread giving rise to elongated fingers at lower *We* in LM. (C) The mode growth time is plotted against *Bo*. The growth time is low for the lower *Bo*, confirms the large amplitude observed at lower *Bo*. (D) Number of fingers variation with *We*. The dotted line represents the theoretical prediction.

The amplitude ($A_f$) is observed to increase with the decrease in the rim Bond number. As rim width ($b$) decreases with an increase in $We$ (Supplementary Fig. S5), higher impact velocities lead to lower $Bo$. Mode growth time, given by $t_g = \tau_r/\omega$ is plotted in Fig. 7(C). $\tau_r$ is evaluated from the experimentally measured rim width ($b$), and $\omega$ is extracted from the theoretical curve shown in Supplementary Fig. S3. Growth time for the fastest mode increases with $Bo$. This explains the larger amplitude at smaller $Bo$. The wavelength of the perturbation scales with maximum spread as $\lambda \sim \pi D_{max}/N_f$, where $N_f$ is the number of fingers and $\lambda$ is the wavelength of the fastest growing mode. Supplementary Fig. S6 plots the theoretical wavelength of the fastest growing mode for different $Bo$. We can assume $\lambda \cong \pi b$. This implies $N_f \sim D_{max}/b$. Assuming most of the volume during spreading transfers to the rim results in $b \sim (D_0^3/D_{max})^{0.5}$. Thus, the number of fingers should scale with $N_f \sim D_{max}/b \sim (D_{max}/D_0)^{3/2} \sim We^{3/8}$. This is in good agreement with the observed number of perturbations at the maximum spread in Fig. 7(D).

## 3.5 Effect of viscous loss

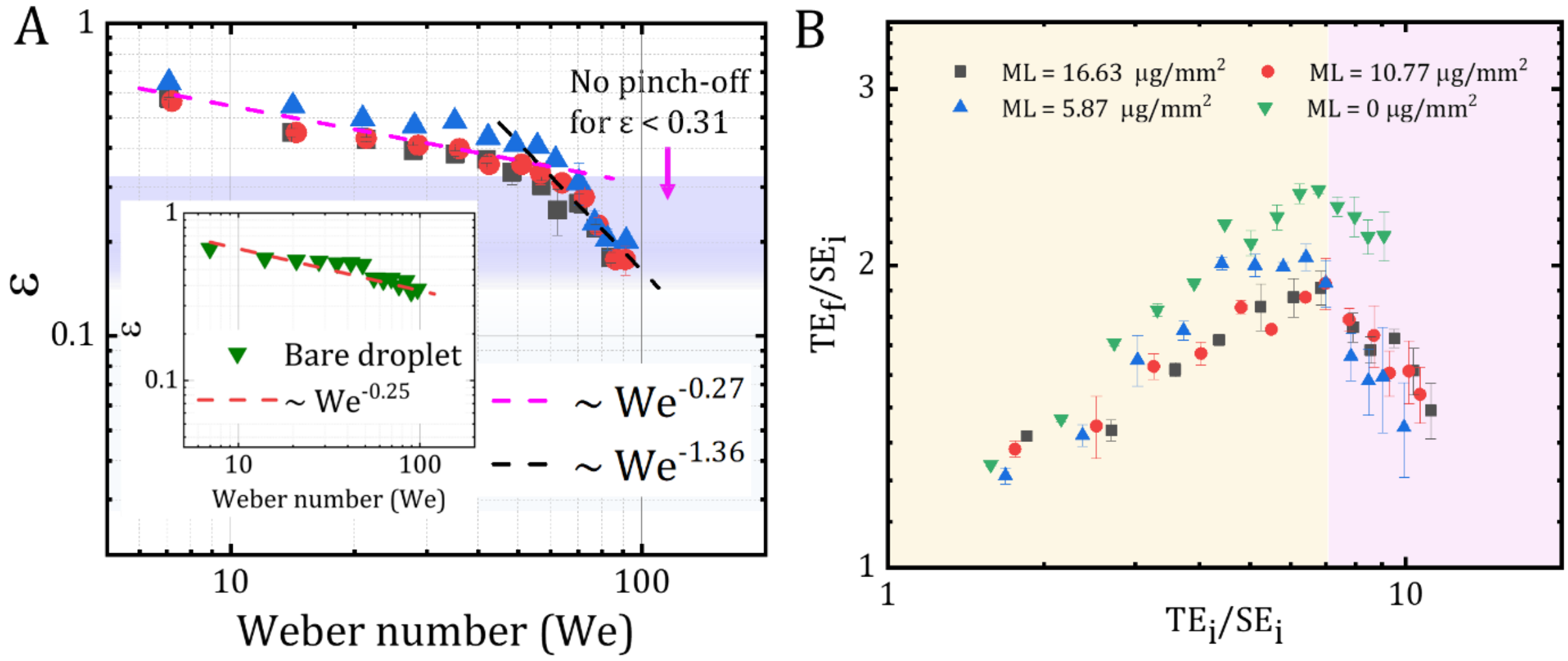

**Fig. 8.** (A) Restitution coefficient of the main droplet. Legend is shown in part (B). The colored region indicates the value of $\varepsilon$ below 0.31. Restitution follows the scaling of slope -0.27 up to $\varepsilon = 0.31$. Consequently, a faster decay in restitution (with a -1.36 slope) is observed (colored region). In contrast, the bare droplet follows the standard scaling of -0.25 up to receding breakup (Inset). (B) Final total energy ($TE_f$) normalized by initial surface energy ($SE_i$) is plotted against normalized total initial energy ($TE_i/SE_i$). The boundary of the colored regions represents the starting of energy loss.

During droplet rebound, the finger formation accounts for additional energy dissipation. To estimate additional losses due to finger collapse, the restitution coefficient $\varepsilon = \left(KE_f/KE_i\right)^{0.5}$ and normalized total energy is plotted in Fig. 8(A) & (B), respectively. $KE_f$ and $KE_i$ are final and initial kinetic energy, respectively. The energy is calculated by tracking the center of mass of the droplet. It is evident from Fig. 8(A) that the recurrence of the no-breakage regime coincides with a faster reduction in the restitution coefficient with $We$. Up to $\varepsilon \sim 0.31$, the LM follows the scaling of $\sim We^{-0.27}$ for the restitution coefficient. This is approximately the same scaling as observed for a bare droplet ($\sim We^{-0.25}$) [34]. However, after that, the restitution coefficient follows a different scaling of $\varepsilon \sim We^{-1.36}$. This proves the existence of additional energy dissipation due to the retraction and collapse of fingers in LM. Whereas the bare droplet continues to follow $\varepsilon \sim We^{-0.25}$ even at higher Weber numbers (Inset, Fig. 8(A)). Similarly, the total final energy loss drops suddenly by a large magnitude in the case of LM after finger formation (Fig. 8(B)).

Normalized energy loss in the pinch-off suppression regime is plotted in Fig. 9(A). We hypothesize that the additional energy loss during the collapse of such fingers depends on the viscous flow in the fingers. The energy loss during retraction mainly depends on the shear stress in the elongated fingers, total numbers ($N_f$) and length of the fingers ($A_f$). The shear stress in the fingers can be estimated as $\mu V_{ret}/D_f$, where $\mu$, $V_{ret}$, $r_f$ are the liquid viscosity, retraction velocity, and finger diameter, respectively. Thus, the energy loss ($KE_i - KE_f$) should scale with $\mu V_{ret} N_f {A_f}^2$. The velocity of retraction depends upon the Taylor-Culick relation: $V_{ret} \sim (\gamma/\rho D_f)^{0.5} \sim V\, We^{-0.25}$, where $V$ is the velocity of impact and $D_f \sim {D_0}^3/{D_{max}}^2$ derived from volume conservation [35]. From the previous scaling arguments, $N_f \sim We^{3/8}$ and $A_f \sim D_{max} \sim D_0\, We^{0.25}$ because of the major growth of fingers during the retraction phase. The initial kinetic energy ($KE_i$) scales as $\rho V^2 D_0^3$. Thus, the normalized loss should scale with

$$\frac{KE_i - KE_f}{KE_i} \sim \frac{We^{5/8}}{Re} \tag{2}$$

where $Re$ is the Reynolds number. Fig. 9(A) shows scaling according to Eq. (2), which agrees very well with the experiments.

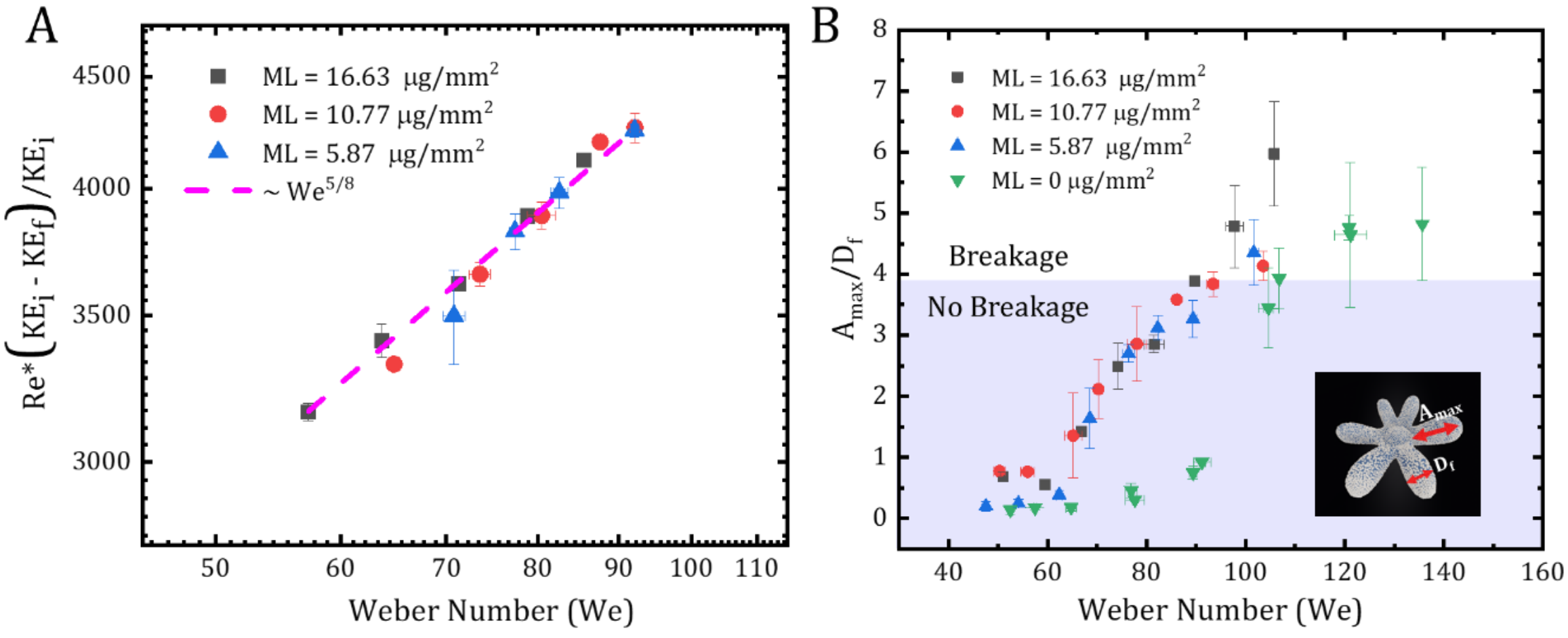

**Fig. 9.** (A) Normalized energy loss plotted against Weber number. The scaling of slope 5/8 matches very well with experimental results. (B) The ratio of maximum finger amplitude ($A_{max}$) to finger diameter ($D_f$) is plotted against the *We*. The colored region indicates below critical $A_{max}/D_f \sim 3.8$, where no breakage of fingers is present. Inset: Schematic representation of maximum finger amplitude ($A_{max}$) to finger diameter ($D_f$).

Despite forming prominent fingers, we do not observe its breakage into smaller droplets. As $A_f \sim D_{max}$, at lower $We$, the resultant finger length is insufficient to destabilize and break into droplets. Finger breakage due to RP instability in jets [36] is possible when the $A_{max}/D_f > \pi$. Fig. 9(B) represents the ratio of maximum finger amplitude ($A_{max}$) and finger diameter ($D_f$). We found the critical ratio of $A_{max}/D_f$ ~ 3.8 is responsible for the breakup of the fingers. Despite prominent fingers in 57 < *We* < 92 regions, the $A_{max}/D_f$ ratio for LM remains lower than the critical limit. Thus, fingers remain stable without dissociation.

### 3.6 Role of the initial perturbation

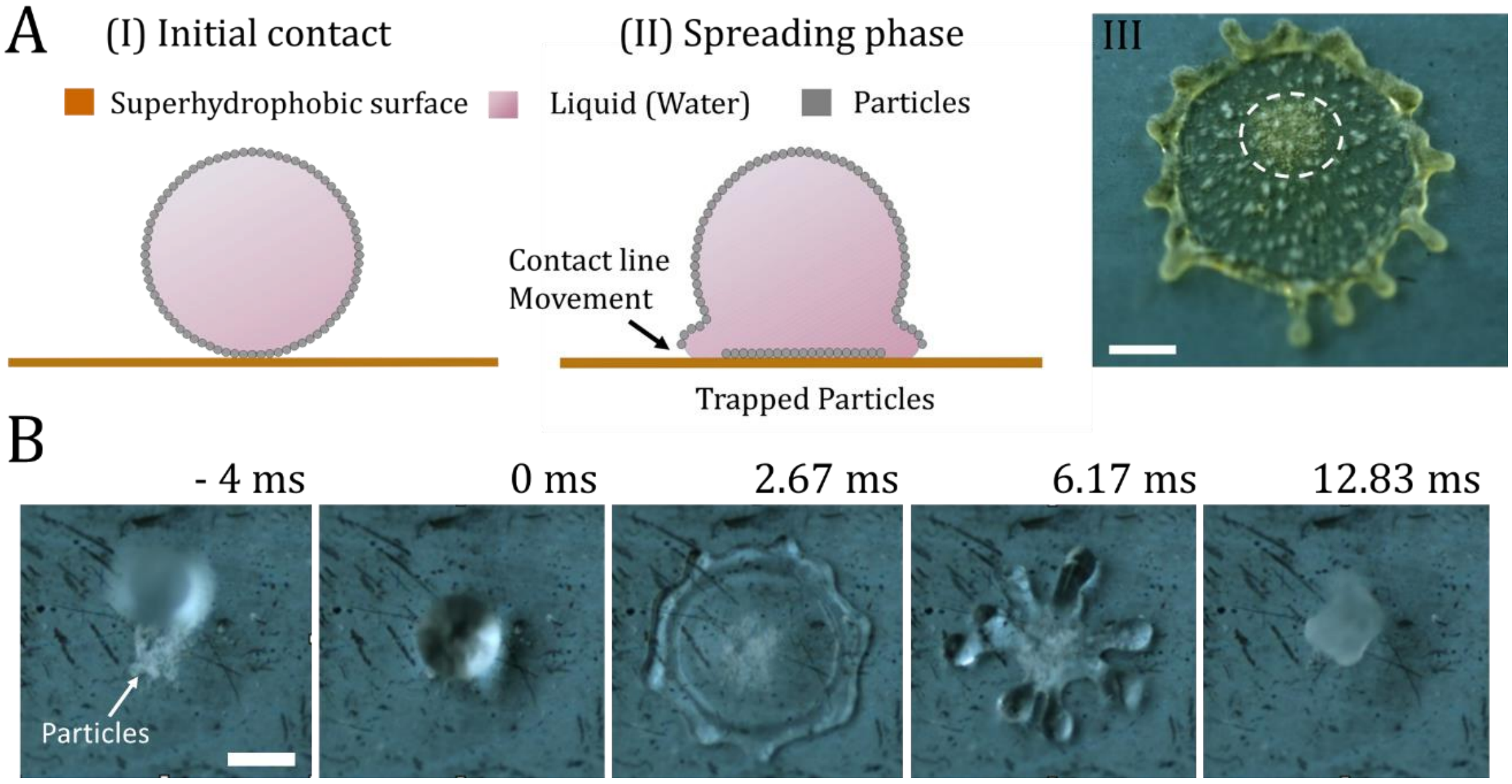

**Fig. 10.** (I) Initial contact of LM with the superhydrophobic surface (II) Movement of contact line during a spreading phase where particles get trapped at the impact position. (III) Trapped particles for ML = 16.63 μg/mm$^2$ Liquid marble at 120 *We*. A slightly tilted surface and higher *We* are used for visualizing the trapped particles clearly. Trapped particles appear to be asymmetric due to the tilted view (scale bar ~ 2 mm). (B) Bare droplet impact over the surface with PTFE particles at *We* ~ 72 shows a similar finger formation, further confirming the role of initial perturbation and particles. Scale = 2 mm.

The growth of an unstable mode is represented by $A_f = A_o exp\,(\beta t)$, where $A_o$ is the initial perturbation, $\beta$ is the mode-specific growth rate, and $t$ is the growth time [37]. Since the same phenomena govern the instability in both LM and bare droplet cases, the growth rate and time should remain the same for a given impact condition. This is verified experimentally in Fig 7(A) & (C). Hence, we postulated that early finger formation in LM is because of the higher initial perturbations happening during the initial phase of impact (< 1 ms). The initial perturbations result from trapped particles. During initial contact, particles get trapped between the spreading liquid and the superhydrophobic surface (Fig. 6(A) and Fig. 10(A)). The trapped particles make the bottom surface inhomogeneous. The contact line experiences wetting discontinuity at the edge of the trapped particle region. A separate experiment was also performed to verify the role of particles trapped between the liquid and the superhydrophobic surface. A small number of particles are pre-deposited on the superhydrophobic surface. It is observed that fingering instability is triggered even for bare droplets at lower impact energies ($We \sim 72$), further confirming the role of initial perturbation (Fig. 10(B) and Supplementary Video S3).

# 4 Conclusion

The coating of hydrophobic particles over droplets remarkably affects the stability of impacting droplets. The LM shows an anomalous stability regime at higher *We* (57 to 90). Interestingly, finger formation, which is usually responsible for droplet fragmentation, is helping LM to be stabilized at higher *We*. This study can have implications in understanding the physics behind several biological phenomena, such as disease spreading in plants through pollen-laden droplets and rupture models of cells and organs. Moreover, such mechanical stability can be applied to design better and continuous Janus LM production and splitting devices. [14,38] Besides, this paper indicates the possible way to increase the mechanical stability of fragile droplet-based chemical and biological reactors by replacing them with LM-based reactors. The phenomenon of early finger formation is not limited to LM impact on superhydrophobic surfaces. A similar finger formation can be seen on heated surfaces above Leidenfrost temperatures (Supplementary Video S4). Such scenarios also help better understand interactions between droplets and dust/pollen particles, making them extremely useful in designing self-cleaning surfaces and pollen-laden interfaces [5,6,19,20].

## Data availability:

Source data are provided in this paper. All other data that support the plots within this paper are available from the corresponding author upon reasonable request.

## Acknowledgments:

The authors would like to acknowledge the National Nanofabrication Centre and the Micro/Nano Characterization Facility at CeNSE, IISc for the fabrication and characterization. The authors also acknowledge Arunsingh Baghel for his help in writing image analysis code. All the authors would like to thank the Department of Science and Technology and Ministry of Education, Government of India for financial support. RL acknowledges Prime Minister's Research Fellowship for the financial support.

## Contributions:

RL, CDM, and PS designed the experiments and wrote the paper. RL and CDM performed the experiments. RL and PS set up the model. PS supervised the project and acquired the funding. All authors have read and approved the final version of the paper.

## References:

[1] A.U. Siddique, M. Trimble, F. Zhao, M.M. Weislogel, H. Tan, Jet ejection following

drop impact on micropillared hydrophilic substrates, Phys. Rev. Fluids. 5 (2020) 063606. https://doi.org/10.1103/PhysRevFluids.5.063606.

[2] D. Bartolo, C. Josserand, D. Bonn, Singular Jets and Bubbles in Drop Impact, Phys. Rev. Lett. 96 (2006) 124501. https://doi.org/10.1103/PhysRevLett.96.124501.

[3] X. Deng, C. Huang, S. Liu, C. Yang, P. Wu, J. He, W. Jiang, Urea Melt Marbles Developed by Enwrapping Urea Melt Droplets with Superhydrophobic Particles: Preparation, Properties, and Application in Large Urea Granule Production, Adv. Mater. Interfaces. 8 (2021) 2100253. https://doi.org/10.1002/ADMI.202100253.

[4] M. Damak, M.N. Hyder, K.K. Varanasi, Enhancing droplet deposition through in-situ precipitation, Nat. Commun. 2016 71. 7 (2016) 1–9. https://doi.org/10.1038/ncomms12560.

[5] S. Kim, H. Park, H.A. Gruszewski, D.G. Schmale, S. Jung, Vortex-induced dispersal of a plant pathogen by raindrop impact, Proc. Natl. Acad. Sci. 116 (2019) 4917–4922. https://doi.org/10.1073/PNAS.1820318116.

[6] S. Nath, S.F. Ahmadi, H.A. Gruszewski, S. Budhiraja, C.E. Bisbano, S. Jung, S.G. III, J.B. Boreyko, 'Sneezing' plants: pathogen transport via jumping-droplet condensation, J. R. Soc. Interface. 16 (2019). https://doi.org/10.1098/RSIF.2019.0243.

[7] D. Roy, S. M, A. Rasheed, P. Kabi, A.S. Roy, R. Shetty, S. Basu, Fluid dynamics of droplet generation from corneal tear film during non-contact tonometry in the context of pathogen transmission, Phys. Fluids. 33 (2021) 092109. https://doi.org/10.1063/5.0061956.

[8] A. Tripathy, G. Muralidharan, A. Pramanik, P. Sen, Single etch fabrication and characterization of robust nanoparticle tipped bi-level superhydrophobic surfaces, RSC Adv. 6 (2016) 81852–81861. https://doi.org/10.1039/c6ra16312b.

[9] Z. Liu, X. Pan, Q. Ma, H. Fang, Receding Dynamics of Droplet Deposition on a Smooth Surface from a Central Jet to Secondary Droplet Emission, Langmuir. 36 (2020) 15082–15093. https://doi.org/10.1021/acs.langmuir.0c02643.

[10] M. Reyssat, A. Pépin, F. Marty, Y. Chen, D. Quéré, Bouncing transitions on microtextured materials, EPL (Europhysics Lett. 74 (2006) 306. https://doi.org/10.1209/EPL/I2005-10523-2.

[11] X. Huang, K.T. Wan, M.E. Taslim, Axisymmetric rim instability of water droplet impact on a superhydrophobic surface, Phys. Fluids. 30 (2018) 094101. https://doi.org/10.1063/1.5039558.

[12] Y. Wang, R. Dandekar, N. Bustos, S. Poulain, L. Bourouiba, Universal Rim Thickness in Unsteady Sheet Fragmentation, Phys. Rev. Lett. 120 (2018) 204503. https://doi.org/10.1103/PHYSREVLETT.120.204503.

[13] M. Ghanbari, G. Rezazadeh, A liquid-state high sensitive accelerometer based on a micro-scale liquid marble, Microsyst. Technol. 26 (2020) 617–623. https://doi.org/10.1007/s00542-019-04528-7.

[14] B. Wang, K.F. Chan, F. Ji, Q. Wang, P.W.Y. Chiu, Z. Guo, L. Zhang, On-Demand Coalescence and Splitting of Liquid Marbles and Their Bioapplications, Adv. Sci. 6 (2019) 1802033. https://doi.org/10.1002/advs.201802033.

[15] P.K. Roy, B.P. Binks, S. Fujii, S. Shoval, E. Bormashenko, S. Fujii, S. Shoval, E. Bormashenko, Composite Liquid Marbles as a Macroscopic Model System Representing Shedding of Enveloped Viruses, J. Phys. Chem. Lett. 11 (2020) 4279–4285. https://doi.org/10.1021/acs.jpclett.0c01230.

[16] E. Jambon-Puillet, T.J. Jones, P.-T.T. Brun, Deformation and bursting of elastic capsules impacting a rigid wall, 16 (2020) 585–589. https://doi.org/10.1038/s41567-

020-0832-x.

[17] C.H. Ooi, R. Vadivelu, J. Jin, K.R. Sreejith, P. Singha, N.-K.N.-T. Nguyen, N.-K.N.-T. Nguyen, Liquid marble-based digital microfluidics – fundamentals and applications, Lab Chip. 21 (2021) 1199–1216. https://doi.org/10.1039/D0LC01290D.

[18] F. Geyer, M. D'Acunzi, A. Sharifi-Aghili, A. Saal, N. Gao, A. Kaltbeitzel, T.F. Sloot, R. Berger, H.J. Butt, D. Vollmer, When and how self-cleaning of superhydrophobic surfaces works, Sci. Adv. 6 (2020). https://doi.org/10.1126/SCIADV.AAW9727.

[19] X. Yan, B. Ji, L. Feng, X. Wang, D. Yang, K.F. Rabbi, Q. Peng, M.J. Hoque, P. Jin, E. Bello, S. Sett, M. Alleyne, D.M. Cropek, N. Miljkovic, Particulate–Droplet Coalescence and Self-Transport on Superhydrophobic Surfaces, ACS Nano. 16 (2022) 12910–12921. https://doi.org/10.1021/ACSNANO.2C05267.

[20] A. Sayyah, M.N. Horenstein, M.K. Mazumder, Energy yield loss caused by dust deposition on photovoltaic panels, Sol. Energy. 107 (2014) 576–604. https://doi.org/10.1016/J.SOLENER.2014.05.030.

[21] E. Bormashenko, Liquid marbles: Properties and applications, Curr. Opin. Colloid Interface Sci. 16 (2011) 266–271. https://doi.org/10.1016/j.cocis.2010.12.002.

[22] R. Wang, X. Li, On the effective surface tension of powder-derived liquid marbles, Powder Technol. 367 (2020) 608–615. https://doi.org/10.1016/j.powtec.2020.04.028.

[23] Y. Asaumi, M. Rey, N. Vogel, Y. Nakamura, S. Fujii, Particle Monolayer-Stabilized Light-Sensitive Liquid Marbles from Polypyrrole-Coated Microparticles, Langmuir. 36 (2020) 2695–2706. https://doi.org/10.1021/acs.langmuir.0c00061.

[24] M. Tenjimbayashi, S. Samitsu, Y. Watanabe, Y. Nakamura, M. Naito, Liquid Marble Patchwork on Super-Repellent Surface, Adv. Funct. Mater. (2021) 2010957.

https://doi.org/10.1002/adfm.202010957.

[25] W. Zhang, X. Wen, S. Yang, Y. Berta, Z.L. Wang, Single-crystalline scroll-type nanotube arrays of copper hydroxide synthesized at room temperature, Adv. Mater. 15 (2003) 822–825. https://doi.org/10.1002/adma.200304840.

[26] A. Kumar, A. Tripathy, C.D. Modak, P. Sen, Designing assembly of meshes having diverse wettability for reducing liquid ejection at terminal velocity droplet impact, J. Microelectromechanical Syst. 27 (2018) 866–873. https://doi.org/10.1109/JMEMS.2018.2850903.

[27] H. Hoffman, R. Sijs, T. De Goede, D. Bonn, Controlling droplet deposition with surfactants, Phys. Rev. Fluids. 6 (2021) 033601. https://doi.org/10.1103/PHYSREVFLUIDS.6.033601.

[28] M. Aytouna, D. Bartolo, G. Wegdam, D. Bonn, S. Rafaï, Impact dynamics of surfactant laden drops: Dynamic surface tension effects, Exp. Fluids. 48 (2010) 49–57. https://doi.org/10.1007/S00348-009-0703-9.

[29] C. Clanet, C. Béguin, D. Richard, D. Quéré, Maximal deformation of an impacting drop, J. Fluid Mech. 517 (2004) 199–208. https://doi.org/10.1017/S0022112004000904.

[30] V. Bergeron, D. Bonn, J.Y. Martin, L. Vovelle, Controlling droplet deposition with polymer additives, Nat. 2000 4056788. 405 (2000) 772–775. https://doi.org/10.1038/35015525.

[31] S. Shakeri, S. Chandra, Splashing of molten tin droplets on a rough steel surface, Int. J. Heat Mass Transf. 45 (2002) 4561–4575. https://doi.org/10.1016/S0017-9310(02)00170-9.

[32] Y. Liu, P. Tan, L. Xu, Kelvin-Helmholtz instability in an ultrathin air film causes drop splashing on smooth surfaces, Proc. Natl. Acad. Sci. U. S. A. 112 (2015) 3280–3284. https://doi.org/10.1073/PNAS.1417718112.

[33] S.S. Yoon, R.A. Jepsen, S.C. James, J. Liu, G. Aguilar, Are Drop-Impact Phenomena Described by Rayleigh-Taylor or Kelvin-Helmholtz Theory?. 27 (2009) 316–321. https://doi.org/10.1080/07373930802682858.

[34] D.G.K. Aboud, A.-M. Kietzig, On the Oblique Impact Dynamics of Drops on Superhydrophobic Surfaces. Part II: Restitution Coefficient and Contact Time, Langmuir. 34 (2018) 9889–9896. https://doi.org/10.1021/ACS.LANGMUIR.8B01233.

[35] Z. Hu, F. Chu, X. Wu, Double-peak characteristic of droplet impact force on superhydrophobic surfaces, Extrem. Mech. Lett. (2022) 101665. https://doi.org/10.1016/J.EML.2022.101665.

[36] AL Yarin, I. V. Roisman, C. Tropea, Collision phenomena in liquids and solids, Collis. Phenom. Liq. Solids. (2017) 1–614. https://doi.org/10.1017/9781316556580.

[37] Y. Wang, L. Bourouiba, Growth and breakup of ligaments in unsteady fragmentation, J. Fluid Mech. 910 (2021) A39. https://doi.org/10.1017/JFM.2020.698.

[38] B.S. Lekshmi, A.S. Yadav, P. Ranganathan, S.N. Varanakkottu, Simple and Continuous Fabrication of Janus Liquid Marbles with Tunable Particle Coverage Based on Controlled Droplet Impact, Langmuir. 36 (2020) 15396–15402. https://doi.org/10.1021/acs.langmuir.0c02988.

# Supplementary information

# Suppression of Droplet Breakage by Early Onset of Interfacial Instability

Rutvik Lathia, Chandantaru Dey Modak, and Prosenjit Sen*

Centre for Nano Science and Engineering, Indian Institute of Science, Bangalore, India, 560012

***Corresponding Author's Email:** prosenjits@iisc.ac.in

**Supplementary Figure S1:**

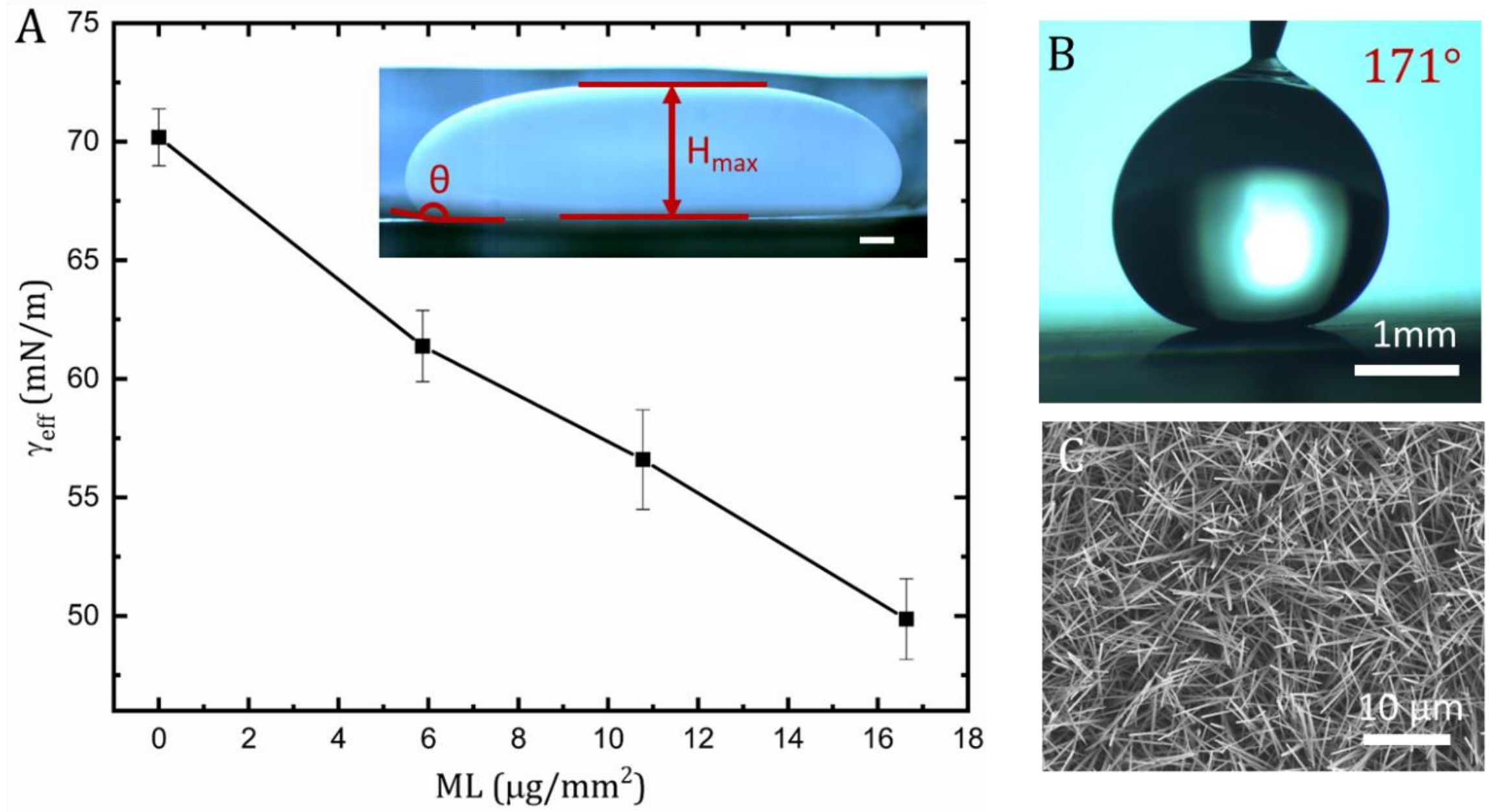

**Supplementary Figure S1:** (A) Change in $\gamma_{eff}$ with an increase in mass loading of the LM. Inset: Maximum puddle height method for determining effective surface tension ($\gamma_{eff}$) of the LM. $H_{max}$ represents the maximum height of the puddle, and $\theta$ is the contact angle of LM with the SHP surface. The volume taken for the effective surface tension measurement is 1000 $\mu L$ (Scale bar - 1 $mm$). (B) Contact angle measurement for the prepared superhydrophobic surface (Scale bar - 1 $mm$). (C) SEM image of the nanowires of superhydrophobic surface (Scale bar – 10 $\mu m$).

**Supplementary Figure S2:**

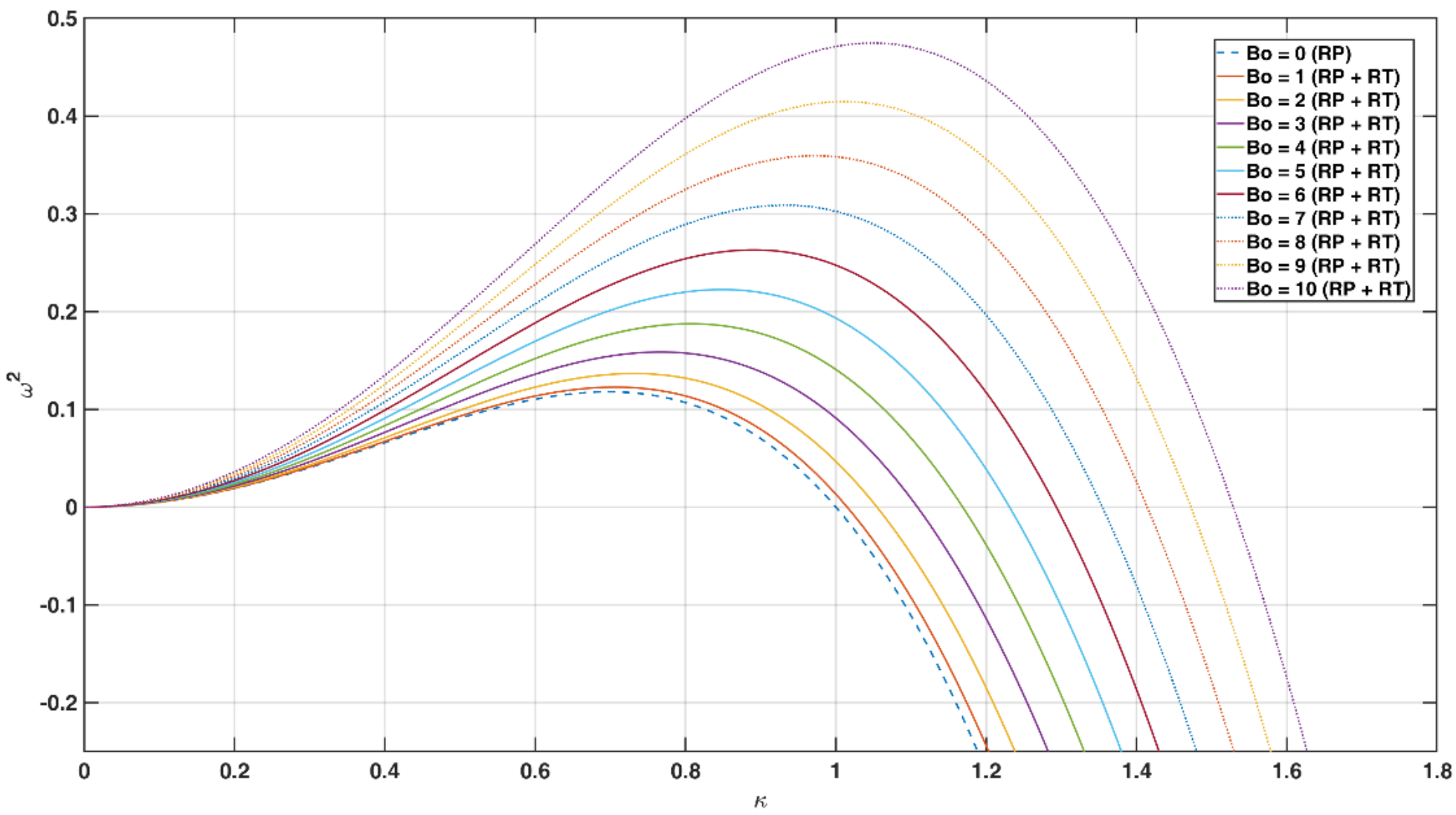

**Supplementary Figure S2:** Dispersion curves for different Bond number. $Bo = 0$ represents the pure RP instability. Increase in $Bo$ results in combined RP and RT instability.

**Supplementary Figure S3:**

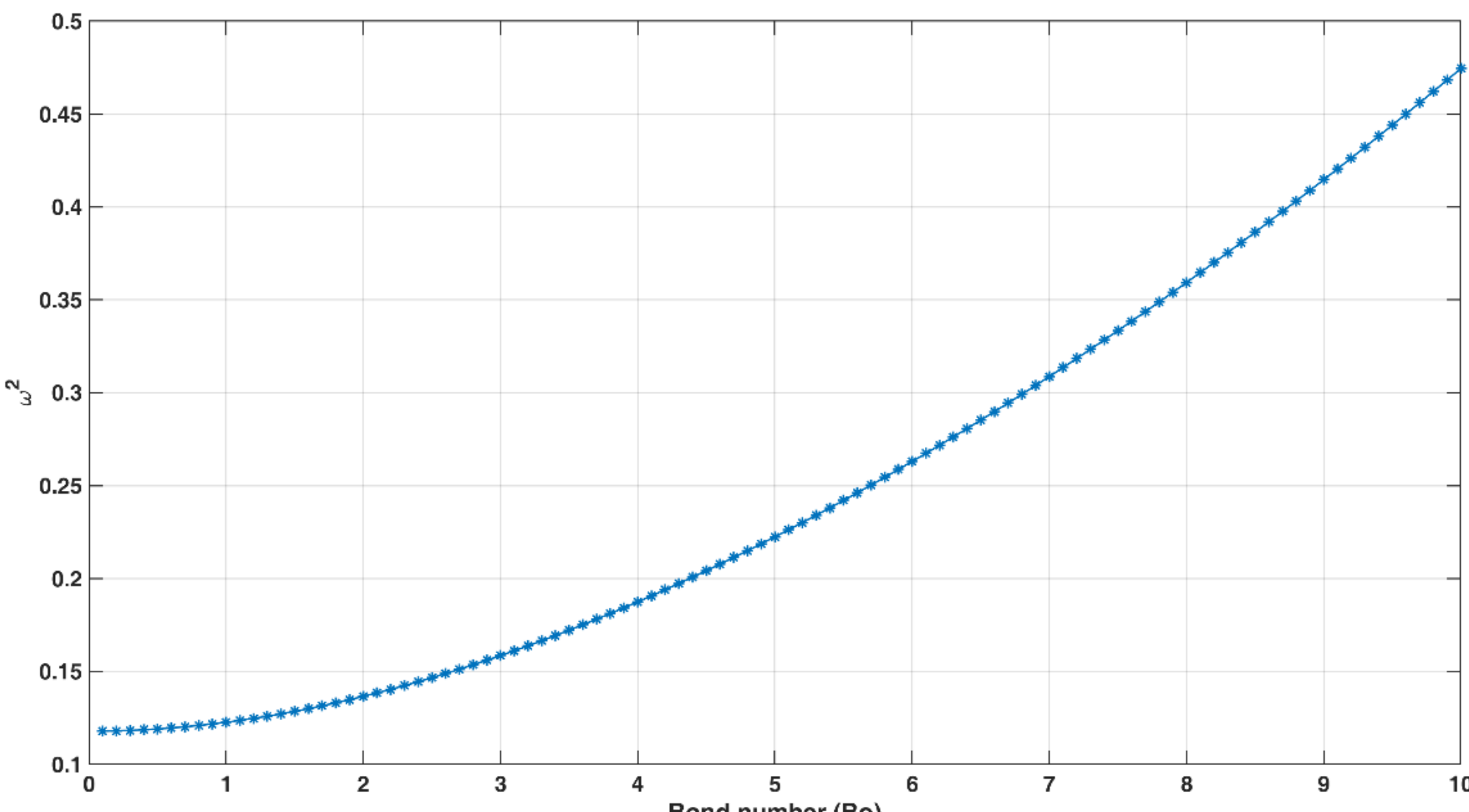

**Supplementary Figure S3:** Growth rate variation for the fastest growing mode with *Bo*.

**Supplementary Figure S4:**

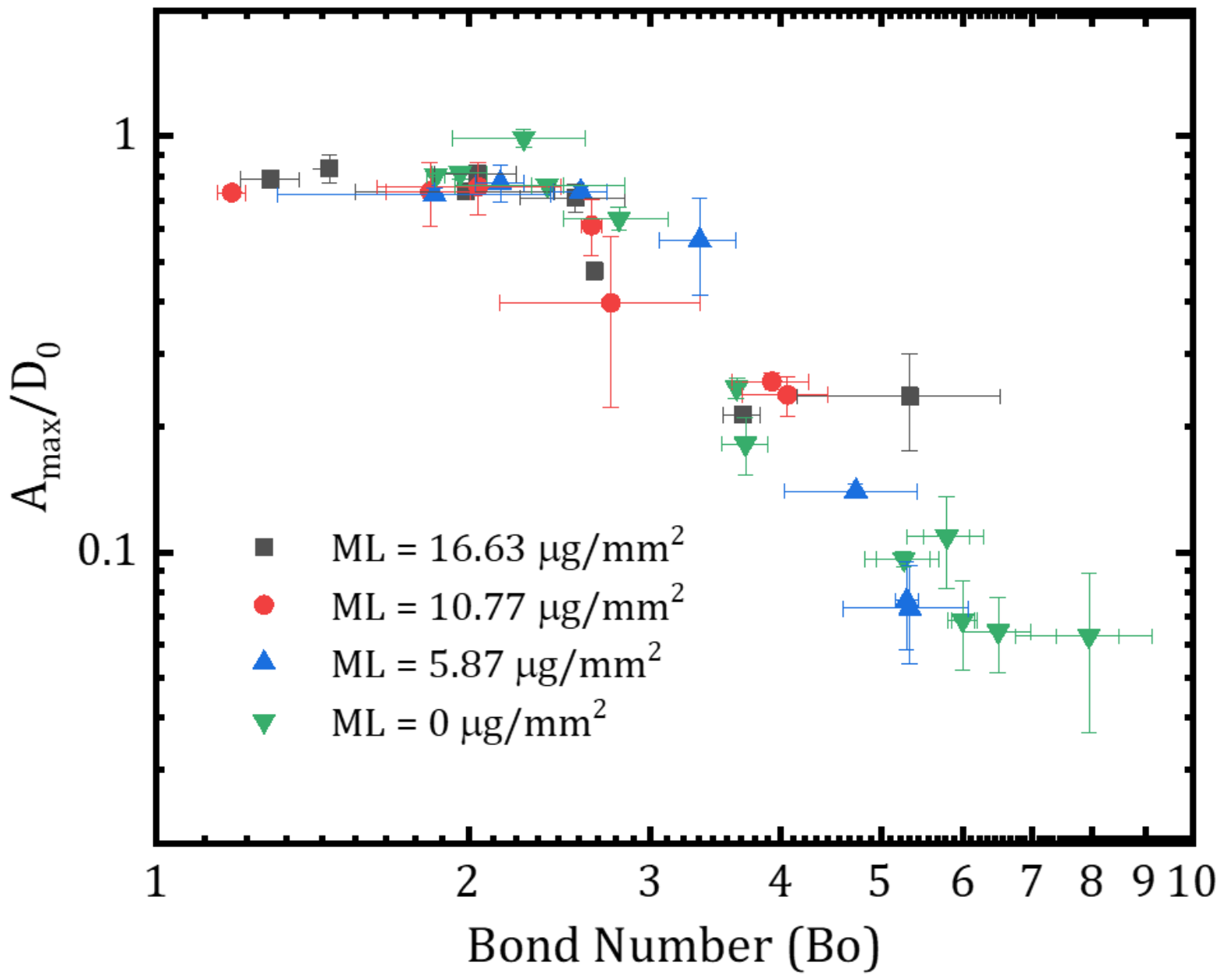

**Supplementary Figure S4:** The normalized maximum amplitude of the finger ($A_{max}/D_0$) is plotted against the rim Bond number at the maximum spread.

**Supplementary Figure S5:**

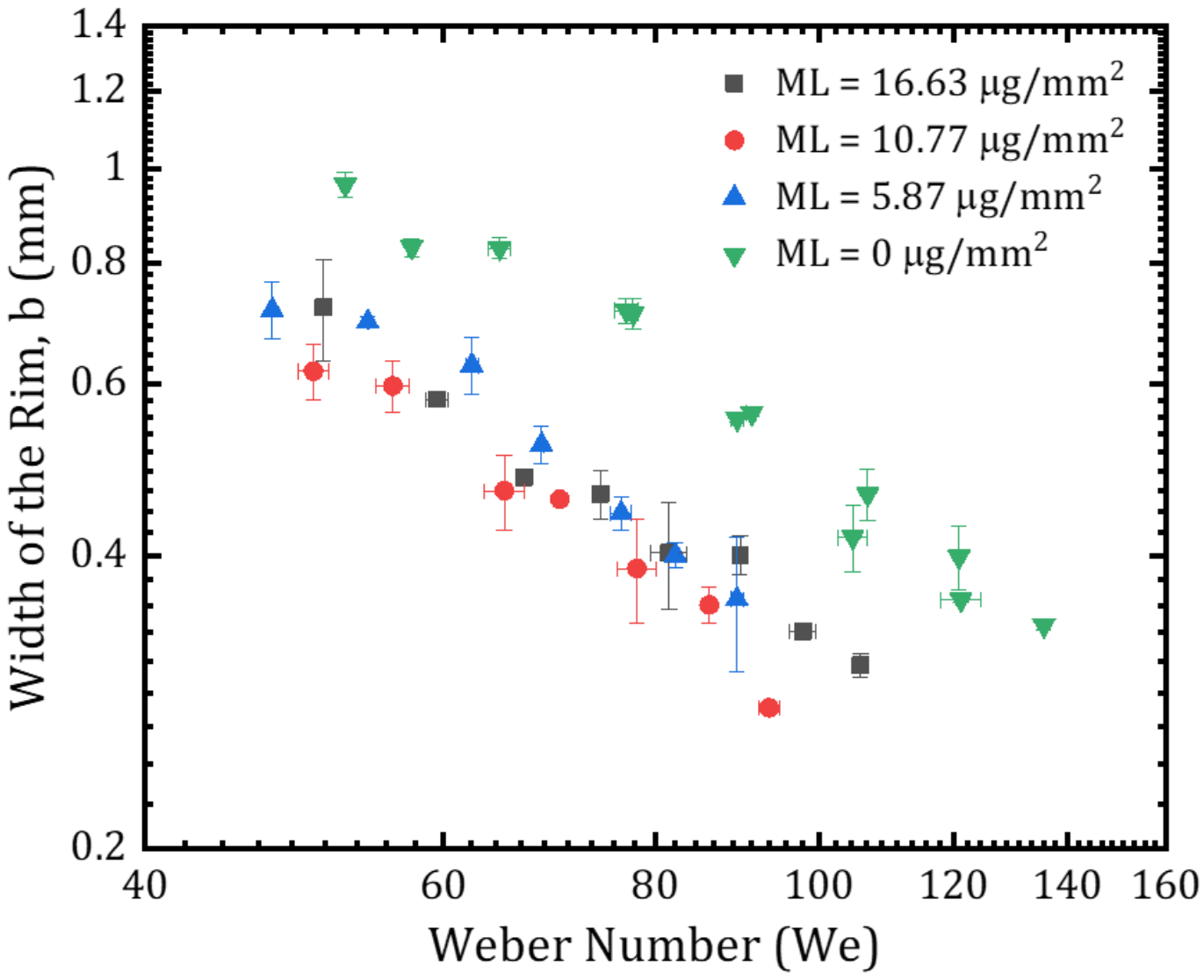

**Supplementary Figure S5:** The rim width variation with *We* for various mass loaded LM. The clear distinction between LM and bare droplet rim can be seen.

**Supplementary Figure S6:**

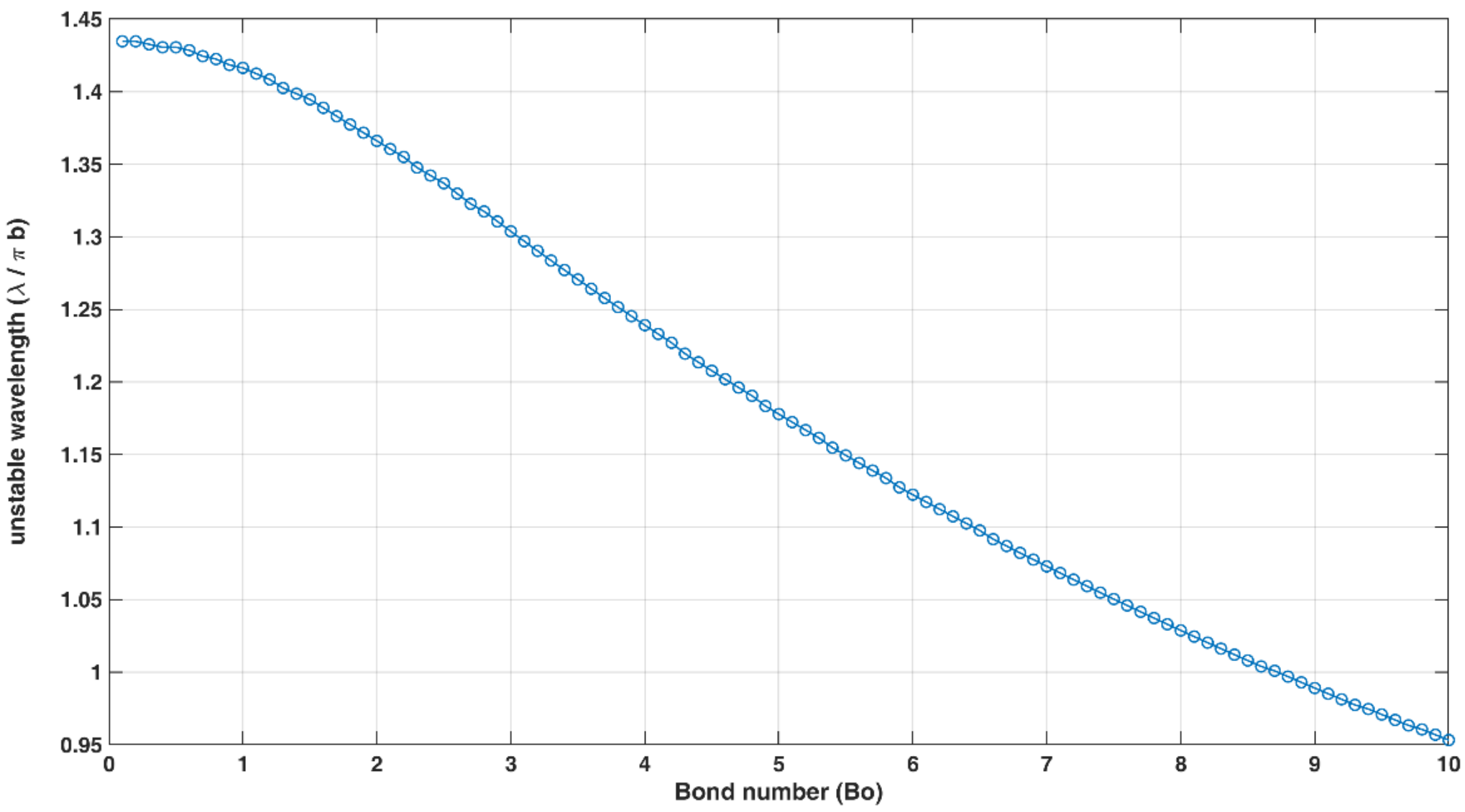

**Supplementary Figure S6:** Wavelength of fastest growing mode for different *Bo*.

## Description of Supplementary Video Files:

**Supplementary Video S1:** Pinch-off behavior of LM with *We*.

**Supplementary Video S2:** Liquid flower formation in LM as compared to the bare drop.

**Supplementary Video S3:** Bare droplet impact over the particle-covered surface.

**Supplementary Video S4:** LM impact on heated (300°C) copper surface.